\documentclass[a4paper,12pt]{article}
\usepackage{amsthm}
\usepackage{amssymb}
\usepackage{amsmath}
\usepackage{amsfonts}
\usepackage{color}
\usepackage{graphicx}
\usepackage[square, comma, sort&compress, numbers]{natbib}

\newtheorem{definition}{Definition}[section]

\def\b1{\mbox{\boldmath $1$}}

\newenvironment{demo*}{\vspace{3mm}\noindent{\bf Proof.}}{\hfill $\Box$ \vspace{3mm}}

\begin{document}
\title{\bf \Large {The $CI$-index: a new index to characterize the scientific output of researchers }}
{\color{red}{\author{
\normalsize{Xuehua Yin,\;\; Xiuyan Sha,\;\; Chuancun Yin}\\
{\normalsize\it  School of Statistics,  Qufu Normal University}\\
\noindent{\normalsize\it Shandong 273165, China}\\
e-mail:  ccyin@qfnu.edu.cn}}}
\maketitle
\vskip0.01cm
\noindent{\large {\bf Abstract}}  { {We propose a simple new index, named the $CI$-index, based on the  Choquet integral to characterize the scientific output of researchers. This index is an improvement of the $A$-index and $R$-index and has a notable feature that
highly cited papers have highly weights and  lowly cited papers have lowly weights. We  proposed a ranking method of   distinguish   researchers  to compare their  scientific outputs. In  applications many researchers may have
the same $h$-index, $g$-index or  $R$-index, however, the $CI$-index can be provided  an effective method of distinguish  among such researchers.}}



\numberwithin{equation}{section}
\section{Introduction}\label{intro}
Since the physicist Hirsch (2005) introduced the so-called $h$-index which as a new  indicator for   measure the impact of a researcher's   scientific  research output, it has got a lot of attention from both in the  scientific community  and the scientometrics (informetrics) literature for its good properties to measure the scientific production of researchers, more than 8430  of articles have been written on the $h$-index
 (Data from Google Scholar as of March 15, 2019). The result was reviewed in Nature (Ball (2005)) and Science (Anon (2005)) as well. A large
part of the literature    building on Hirsch's work is concerned
with introducing variants, extensions, and generalizations of
the $h$-index. In the study of Bornmann et al. (2011), no less than 37 variants of the $h$-index were
listed. In a more recent study (Bornmann (2014)) says that there are  around 50  variants of the $h$-index.
The $h$-index is a simple single number incorporating both  quantitative and   qualitative  aspects. The $h$-index is also robust in the sense that it is insensitive to a set of uncited (or lowly cited) papers but also it is insensitive to one or
several outstandingly highly cited papers. This last aspect can be considered as a
drawback of the $h$-index, for more details we refer to Egghe (2006a).
For more advantages and disadvantages of the $h$-index see also Hirsch (2005, 2007) and Jin et al.
(2007). In order to overcome some of these  limitations  scientists have proposed several  new indicators based on the $h$-index
 with the intention of either replacing or complementing the original $h$-index.  To
overcome the well-known problem of the insensitivity of the h-index  to the
number of citations received by highly cited paper,
Egghe (2006a, 2006b)
first developed the $g$-index, Jin (2006) followed with the $A$-index,  Jin et al. (2007) suggested correcting the $h$-index for the aging of papers using the $AR$-index, Egghe and  Rousseau (2008) proposed a
citation-weighted $h$-index, Anderson et al. (2008) describe a new version of the $h$-index, the ``tapered h-index", which
positively scores all of an author's citations, accounting for
the tapered distribution of citations associated with highly cited papers rather than using
a cut-off at $h$.   Hirsch (2010) proposed an index to quantify an individual's scientific research
output that takes into account the effect of multiple
coauthorship. Alonso et al. (2010) proposed  a new index, called $hg$-index, to characterize the
scientific output of researchers which is based on both $h$-index and $g$-index to overcome the limitations of both indices.
Recall that $A$-index is simply defined as the average number of citations received by the publications included
in the Hirsch core. Mathematically, this is
$A=h^{-1}\sum_{i=1}^h y_i,$
where the numbers of citations $y_i$'s are ranked in decreasing order. Note that all the  $y_i$ have the same weight $\frac{1}{h}$.
A closer related to $h$-index is $R$-index introduced in Jin et al. (2007) which is defined as $R=\sqrt{h A}$.
Recently, Perry and Reny (2016) introduced   Euclidean index $l_E$, which is the Euclidean length of $(x_1,\cdots,x_n)$. Note that the Euclidean index, to some extent, avoids several shortcomings of the $h$-index and its successors.   The only drawback may be that the weight of highly cited papers is too high.  For example, consider three researchers $A$, $B$ and $C$, $A$ has 100 papers, each has 10 citations ($h$= 10),
 $B$ has 1 paper   with 100  citations ($h$= 1) and   $C$ has 10000 papers, each  has 1 citation ($h$= 1). Then  $A$, $B$ and $C$ have the same  Euclidean index 100.

The aim of this paper is to present a new index---called the  Choquet integral  index ($CI$-index for short)- to characterize the
scientific output of researchers. This index is an improvement of the $A$-index and has a notable feature that
highly cited papers have highly weights and  lowly cited papers have lowly weights. To our best knowledge, such  a index has not been studied  before.
In the following,  we first  recall the definitions of distortion  function and distortion expectation or  Choquet integral, then we introduce three $CI$-indices, namely,
$CI_h$-index in the $h$-core,  $CI_{g}$-index in the $g$-core and $CI_{N}$-index in  the $N$-core, where $N$ stands for   all citations. After that we put forward a method to compare two researchers.

\vskip 0.2cm
\numberwithin{equation}{section}
\section {Preliminaries}
\begin{definition}
 A vector $W=(w_1,\cdots, w_n)$ is called a  weight if $W$ having the properties
$$w_1+\cdots+w_n=1,\;\; 0\le w_i\le 1, i=1,2,\cdots,n.$$
Moreover, if $w_i\le w_j \; \forall \; i<j$, then we call $W$ is increasing, on the other hand if $w_i\ge w_j \; \forall \; i<j$, we call it is
decreasing.
\end{definition}
\begin{definition} A distortion function is a non-decreasing function $g:[0,1]\rightarrow [0,1]$ such that
$g (0) = 0, g (1) =1$.
\end{definition}
The  notion of distortion function was proposed by  Yaari (1987)  in dual
theory of choice under risk, since then many different distortions $g$ have been proposed in
the literature.  The distortion function is also called  regular increasing monotone quantifier in computer science and  artificial intelligence literature, see Yager (1996).  Here we list some commonly used distortion functions:\\
 $\bullet$ Incomplete beta function $g(x)=\frac{1}{\beta(a,b)}\int^x_0t^{a-1}(1-t)^{b-1}dt$, where $a>0$ and $b>0$ are parameters and
 $\beta(a,b)=\int^1_0 t^{a-1}(1-t)^{b-1}dt$. Setting $b=1$ gives the power distortion $g(x)= x^{a}$; setting $a=1$
gives the dual-power distortion $g(x)= 1-(1-x)^{b}.$\\
$\bullet$  The Wang distortion $g(x)=\Phi(\Phi^{-1}(x)+\Phi^{-1}(p)), 0<p<1,$
where $\Phi$ is the distribution function of the standard normal.\\
$\bullet$ The  lookback distortion  $g(x)=x^p(1- p \ln x),  p\in (0, 1].$

Let us recall the standard definitions of convexity and  concavity of functions.
  \begin{definition} Let $I$ be an interval in real line $\Bbb{R}$. Then the function $f: I\rightarrow\Bbb{R}$ is said to be convex if
for all $x, y \in I$ and all $\alpha\in [0,1]$, the inequality
$$f(\alpha x+(1-\alpha)y)\le \alpha f(x)+(1-\alpha)f(y)$$
holds. If this inequality is strict for all $x\neq y$ and $\alpha\in (0,1)$, then $f$ is said to be strictly convex.
A closely related concept is that of concavity: $f$ is said to be (strictly) concave if,
and only if, $-f$ is (strictly) convex.
\end{definition}
 Assume that $x_1\ge x_2\ge \cdots \ge x_n$ are $n$ positive numbers, given a  distortion  function  $Q$, considering the following weighted sum
\begin{equation}
WS_Q[X]=w_1x_1+\cdots w_n x_n,
\end{equation}
where   $w_i$ is the  weights generated  by  $Q$ as follows
$$w_i=Q\left(1-\frac{i-1}{n}\right)-Q\left(1-\frac{i}{n}\right), i=1,2,\cdots,n.$$
Because of the nondecreasing nature of $Q$ it follows that $w_i>0, i=1,2,\cdots,n$. Furthermore,
from $Q( 1) = 1$ and $Q(0) = 0$, it follows that $\sum_{i=1}^n w_i = 1$.
If $Q$ is convex, then $\{{w}_i\}$ is monotonic  decreasing; if $Q$ is concave, then $\{{w}_i\}$ is monotonic  increasing (see, e.g., Sha et al. (2018) for details).
We remark that (2.1) can be written as a Choquet integral (see, e.g. Denneberg (1994))  of a random variable $X$  with probability distribution
$P(X=x_i)=\frac{1}{n}, i=1,2,\cdots, n$:
\begin{equation}
 WS_Q[X]=\int_0^{+\infty}Q(S_X(x))dx,
\end{equation}
where $S$ is the decumulative distribution function of  $X$   with probability distribution
$P(X=x_i)=\frac{1}{n}, i=1,2,\cdots, n$. For example,
\begin{eqnarray*}
S_X(x)=\left\{  \begin{array}{lll} 1,  \ &{\rm if}\ x<x_n,\\
1-\frac{1}{n}, \ &{\rm if}\ x_n\le x<x_{n-1},\\
1-\frac{2}{n}, \ &{\rm if}\ x_{n-1}\le x<x_{n-2},\\
\vdots\ & \vdots\\
1-\frac{n-1}{n}, \ &{\rm if}\ x_{2}\le x<x_{1},\\
0,\ &{\rm if}\ x\ge x_1,\\
 \end{array}
  \right.
\end{eqnarray*}
for the case of  $x_1>x_2>\cdots > x_n$, and
\begin{eqnarray*}
S_X(x)=\left\{  \begin{array}{ll} 1,  \ &{\rm if}\ x<x_1,\\
0,\ &{\rm if}\ x\ge x_1,\\
 \end{array}
  \right.
\end{eqnarray*}
for the case of  $x_1=x_2=\cdots=x_n$.

Obviously,    the identity function is the smallest concave distortion function and also the  largest convex distortion function;
Any concave distortion function $Q$ gives more weight to the tail than the identity function $Q(x)=x$, whereas any convex distortion function $Q$ gives less weight to the tail than the identity function $Q(x)=x$.
 If $Q(x)=x$, then $\rho_{Q}[X]=E[X]$, the expectation of $X$.
 If $Q$ is concave, then
\begin{eqnarray}
\rho_{Q}[X]\ge\int_0^{+\infty}S_X(x)dx=E[X],
\end{eqnarray}
and if $Q$ is convex, then
$$\rho_{Q}[X]\le\int_0^{+\infty}S_X(x)dx=E[X].$$
Clearly, if  $Q_1(x)\le Q_2(x)$ for $x\in[0,1]$, then $\rho_{Q_1}[X]\le \rho_{Q_2}[X]$ for any random variable $X$.

From (2.2) we see that the  Choquet integral satisfy the following properties:

a)  Positive homogeneity: $\rho_{Q}[aX]=a\rho_{Q}[X]$ for any non-negative constant $a$;

b) Translation invariance: $\rho_{Q}[X+b]=\rho_{Q}[X]+b$ for any constant $b$;

c)  Monotonicity: $\rho_{Q}[X]\le \rho_{Q}[Y]$ for any two random variables $X$ and $Y$, where $X\le Y$ with probability 1.

  \vskip 0.2cm
 \section{ The indices for different datasets}
\setcounter{equation}{0}

 \vskip 0.1cm
 \subsection{ $CI$-index in the $h$-core}
\setcounter{equation}{0}

A paper belongs to the $h$-core of a scientist if it has $\ge h$ citations  (Hirsch 2010). Hence the $h$-core may contain more than $h$ elements and $h$-core contains exactly $h$ elements if only one paper has  $h$ citations. We will use $C_h$ standard for the set of  $h$-core and the number of elements in  $C_h$ is denoted by  $\sharp C_h$. Note that $C_h$ is a  multiset, which, unlike a set, allows for multiple instances for each of its elements.  The number or cardinality of a multiset is constructed by summing up the multiplicities of all its elements. For example, $C(h)=[1,1,6]$, the element 1 has multiplicity 2,  6  has multiplicity 1 and $\sharp C_h=3$.


Let $y_1,\cdots, y_{\sharp C_h}$ be the elements of $h$-core $C_h$ which are ranked in decreasing order, where $h$ is the $h$-index. Note that
 $y_h=\cdots=y_{\sharp C_h}$. The $CI$-index in the $h$-core
is defined as
\begin{eqnarray}
CI_h[Q]=\sqrt{\sharp C_h[y_1 w_1+\cdots+y_{\sharp C_h} w_{\sharp C_h}]},
\end{eqnarray}
where
$$w_j=Q\left(\frac{j}{\sharp C_h}\right)-Q\left(\frac{j-1}{\sharp C_h}\right),\; j=1,2,\cdots, \sharp C_h.$$
In particular, when $\sharp C_h=h$,
\begin{eqnarray}
CI_h[Q]=\sqrt{h[y_1 w_1+\cdots+y_h w_h]},
\end{eqnarray}
where
$$w_j=Q\left(\frac{j}{h}\right)-Q\left(\frac{j-1}{h}\right),\; j=1,2,\cdots,h.$$
Here   $Q$ is a distortion function. The reason that taking the root  is to prevent the  number  being too large. The distinguishability of $CI_h[Q]$
and $(CI_h[Q])^2$ are same, since the function $f(x)=\sqrt{x}$ is strictly increasing.


 If $Q$ is a concave distortion function, then by (2.3) we get
$CI_h[Q]\ge R$, where $R$ is the $R$-index which defined by $R=\sqrt{\sum_{i=1}^h y_j}$ (see e.g. Jin et al. (2007)).
Taking $Q(x)=x$ in (3.1) yields
$$R_m:=CI_h[Q]=\sqrt{\sum_{i=1}^{\sharp C_h}y_j},$$
which can be seen as the modified version of $R$-index. The $R_m$-index are well-defined   regardless of  $\sharp C_h=h$ or not.

 If $Q(x)=\sqrt{x}$, then
$$\sharp C_h w_j=\sharp C_h\left(\sqrt{\frac{j}{\sharp C_h}}-\sqrt{\frac{j-1}{\sharp C_h}}\right)=\sqrt{\sharp C_h}\left(\sqrt{j}-\sqrt{j-1}\right), \; j=1,2,\cdots, \sharp C_h. $$
Therefore,
\begin{eqnarray}
CI_h[Q]=\sqrt{\sqrt{\sharp C_h}\sum_{j=1}^{\sharp C_h} \left(\sqrt{j}-\sqrt{j-1}\right)y_j}.
\end{eqnarray}
In particular, if $\sharp C_h=h$, then
\begin{eqnarray}
CI_h[Q]=\sqrt{\sqrt{h}\sum_{j=1}^{h} \left(\sqrt{j}-\sqrt{j-1}\right)y_j}.
\end{eqnarray}

\vskip 0.2cm
 \subsection{$CI$-index in the $g$-core}

In order to give more weight to highly
cited articles, Egghe (2006b) proposed the $g$-index.
  The $g$-index was presented by Egghe (2006a, b) as a simple variant of the $h$-index.   A set of papers has a $g$-index $g$ if $g$ is the highest rank such that the top $g$ papers have, together, at least $g^2$ citations.
This also means that the top $g+1$ papers have
less than $(g+1)^2$ cites. Egghe and Rousseau (2008) pointed out that a small variant of the $g$-index is possible by not limiting it to $g\le T$, where $T$ stands for total number of papers. This means that, in these cases, fictitious articles with 0 citations have to be
added. We will use $C_g$ stands for the set of  $g$-core and the number of elements in  $C_g$ is denoted by  $\sharp C_g$. Obviously,  $\sharp C_g=g$.

Let $y_1,\cdots, y_{g}$ be the elements of $g$-core $C_g$ which are ranked in decreasing order, where $g$ is the $g$-index.  One can define an analogous quantity of $CI$-index in the $g$-core
\begin{eqnarray}
CI_g[Q]=\sqrt{g(y_1 u_1+\cdots+y_{g} u_{g})},
\end{eqnarray}
where
$$u_j=Q\left(\frac{j}{g}\right)-Q\left(\frac{j-1}{g}\right),\; j=1,2,\cdots,g.$$  Here $Q$ is a distortion function.

In particular, if  $y_1=\cdots =y_{g}$, then $CI_g[Q]=\sqrt{gy_1}$.  If $Q$ is a concave distortion function, then   by (2.3) we get
$ CI_g[Q]\ge R_g,$ where   $R_g=\sqrt{\sum_{i=1}^g y_j}$,  which is closely related to $A_g$-index (see, Schreiber (2010)).
Taking $Q(x)=x$ in (3.3) yields $CI_g[Q]=R_g$.

 If $Q(x)=\sqrt{x}$, then
\begin{eqnarray}
CI_g[Q]=\sqrt{ \sqrt{g}\sum_{j=1}^g \left(\sqrt{j}-\sqrt{j-1}\right)y_j}.
\end{eqnarray}

\vskip 0.2cm
 \subsection{$CI$-index in the core of all  citations}
\setcounter{equation}{0}

Highly cited papers are, of course, important for the determination of the values  of $CI_h$-index and $CI_g$-index. However, it is  not to take into account the ``tail" papers (with low number of citations).
  Thus, maybe many citations that accompany the most highly cited papers effectively contribute zero. A bibliometric measure of publication output should be  assign a positive score to each new citation as it occurs. It is necessary to consider  the $CI$-index in the core of all  citations.

Let $n$ denote the number of published articles by a scientist, and let $x_i,
i = 1, 2,\cdots, n$, denote the number of citations of the $i$-th most cited article, so that $x_1\ge x_2\ge \cdots \ge x_n>0$. Assume that $N=x_1+ x_2+ \cdots + x_n$ represents the total number of citations received.
The $CI$-index in the   core of all  citations
is defined as
\begin{eqnarray}
CI_N[Q] =\sqrt{N(x_1 v_1+\cdots+x_N v_N)},
\end{eqnarray}
where
$$v_j=Q\left(\frac{j}{N}\right)-Q\left(\frac{j-1}{N}\right),\; j=1,2,\cdots,N.$$ Here $Q$ is a distortion function.
If $Q$ is a concave distortion function, then by (2.3) $CI_N[Q]\ge R_N,$ where $R_N=\sqrt{\sum_{i=1}^N x_i}$.

If $Q(x)=\sqrt{x}$, then
\begin{eqnarray}
CI_N[Q]=\sqrt{\sqrt{N} \sum_{j=1}^N  \left(\sqrt{j}-\sqrt{j-1}\right)x_j}.
\end{eqnarray}

 \vskip 0.2cm
 \section{Ranking method of   distinguish   researchers }
\setcounter{equation}{0}

Note that $CI_h[Q], CI_g[Q]$ and $CI_N[Q]$  are the increasing functions of $h, g, N$ and  $y_i's$.  Thus a method can be proposed to compare the scientific outputs of  two researchers $R_1$ and $R_2$ as follows: Assume   the researcher $R_i$ have $h$-indices  $h_i$, $g$-indices  $g_i$ and the
 total number of citations $N_i$, respectively.

(1)\; If  $CI_{h_1}[Q]>CI_{h_2}[Q]$, then  $R_1 \succ R_2$;

(2)\;  $CI_{h_1}[Q]<CI_{h_2}[Q]$, then  $R_1 \prec R_2$;

(3)\; $CI_{h_1}[Q]=CI_{h_2}[Q]$ and $CI_{g_1}[Q]>CI_{g_2}[Q]$, then  $R_1 \succ R_2$;

(4)\; $CI_{h_1}[Q]=CI_{h_2}[Q]$ and $CI_{g_1}[Q]<CI_{g_2}[Q]$, then  $R_1 \prec R_2$;

(5)\; $CI_{h_1}[Q]=CI_{h_2}[Q]$, $CI_{g_1}[Q]=CI_{g_2}[Q]$ and  $CI_{N_1}[Q]>CI_{N_2}[Q]$, then  $R_1\succ R_2$;

(6)\; $CI_{h_1}[Q]=CI_{h_2}[Q]$, $CI_{g_1}[Q]=CI_{g_2}[Q]$ and  $CI_{N_1}[Q]<CI_{N_2}[Q]$, then  $R_1\prec R_2$;

(7)\; $CI_{h_1}[Q]=CI_{h_2}[Q]$, $CI_{g_1}[Q]=CI_{g_2}[Q]$ and  $CI_{N_1}[Q]=CI_{N_2}[Q]$, then  $R_1\sim R_2$;

where $\prec$ means ``is ranked worse than",   $\succ$ means ``is ranked good than" and $\sim$  means ``is equivalent to".

In order to understand this easily, below we will give two   toy examples.

\noindent {\bf Example 4.1.}\ Suppose that $Q(x)=\sqrt {x}$.\\
(1)\ If $y_1=y_2=50, y_3=3, y_4=1$, then $h=3$, $\sharp C_h=3$ and
$$CI_h=\sqrt{\sqrt{3}\times\left[50+\left(\sqrt{2}-1\right)\times 50+\left(\sqrt{3}-\sqrt{2}\right)\times 3\right]}=11.14;$$
(2)\ If $y_1=y_2=50, y_3=3, y_4=3, y_5=1$, then $h=3$,  $\sharp C_h=4$ and
$$CI_h=\sqrt{\sqrt{4}\times[50+(\sqrt{2}-1)\times 50+(\sqrt{3}-\sqrt{2})\times 3+ (\sqrt{4}-\sqrt{3})\times 3]}=12.04;$$
(3)\ If $y_1=70, y_2=30,y_3=3, y_4=1$, then   $h=3$, $\sharp C_h=3$ and
$$CI_h=\sqrt{\sqrt{3}\times [70+(\sqrt{2}-1)\times 30+(\sqrt{3}-\sqrt{2})\times 3]}=12.02;$$
(4)\ If $y_1=70, y_2=30,y_3=3, y_4=3, y_5=1$, then   $h=3$, $\sharp C_h=4$ and
$$CI_h=\sqrt{\sqrt{4}\times [70+(\sqrt{2}-1)\times 30+(\sqrt{3}-\sqrt{2})\times 3+ (\sqrt{4}-\sqrt{3})\times 3]}=12.98;$$
(5)\ If $y_1=90, y_2=10, y_3=3, y_4=1$, then  $h=3$,   $\sharp C_h=3$ and
$$CI_h=\sqrt{\sqrt{3}\times \left[90+\left(\sqrt{2}-1\right)\times 10+\left(\sqrt{3}-\sqrt{2}\right)\times 3\right]}=12.83;$$
(6)\ If $y_1=90, y_2=10, y_3=3, y_4=3, y_5=1$, then   $h=3$,  $\sharp C_h=4$ and
$$CI_h=\sqrt{\sqrt{4}\times \left[90+\left(\sqrt{2}-1\right)\times 10+\left(\sqrt{3}-\sqrt{2}\right)\times 3+\left(\sqrt{4}-\sqrt{3}\right)\times 3\right]}=13.85;$$
(7)\ If $y_1=40, y_2=30, y_3=20, y_4=13, y_5=4$, then   $h=4$,  $\sharp C_h=4$ and
$$CI_h=\sqrt{\sqrt{4}\times \left[40+\left(\sqrt{2}-1\right)\times 30+\left(\sqrt{3}-\sqrt{2}\right)\times 20+\left(\sqrt{4}-\sqrt{3}\right)\times 13\right]}=11.16;$$
(8)\ If $y_1=\cdots=y_{10}=10, y_{11}=3$, then   $h=10$,  $\sharp C_h=10$ and
$CI_h=\sqrt{10\times 10}=10;$\\
(9)\ If $y_1=100, y_2=3, y_2=1$, then   $h=2$,  $\sharp C_h=2$ and
$$CI_h=\sqrt{\sqrt{2}\times \left[100+\left(\sqrt{2}-1\right)\times 3\right]}=11.97;$$
(10)\ If $y_1=103, y_2=1$, then   $h=1$,  $\sharp C_h=1$ and $CI_h=\sqrt{103}=10.15.$\\
Note that the cases (1),(3) and (5) have the same $h=3$, $\sharp C_h=3$ and  $R=\sqrt{103}$, but  with different $CI_h$-index;  the cases (2),(4) and (6) have the same $h=3$, $\sharp C_h=4$ and  $R_m=\sqrt{106}$, but  with different $CI_h$-indices.  Finally, we get that
$$R_8\prec R_{10}\prec R_1\prec R_7\prec R_9\prec R_3\prec R_2\prec R_5\prec R_4\prec R_6.$$
\noindent {\bf Example 4.2.}\ Suppose that $Q(x)=\sqrt {x}$.\\
(1)\ If $y_1=\cdots=y_{10}=10, y_{11}=0$, then   $h=10$, $\sharp C_h=10$,  $g=10$ and
$$CI_h=\sqrt{10\times 10}=10,\; CI_g=\sqrt{10\times 10}=10;$$
(2)\ If $y_1=\cdots=y_{4}=25, y_{5}=0$, then   $h=4$, $\sharp C_h=4$, $g=10$ and
 $$CI_h=\sqrt{\sqrt{4}\times \left[25+\left(\sqrt{2}-1\right)\times 25+\left(\sqrt{3}-\sqrt{2}\right)\times 25+\left(\sqrt{4}-\sqrt{3}\right)\times 25\right]}=10;$$
 $$CI_g=\sqrt{\sqrt{10}\times \left[25+\left(\sqrt{2}-1\right)\times 25+\left(\sqrt{3}-\sqrt{2}\right)\times 25+\left(\sqrt{4}-\sqrt{3}\right)\times 25\right]}=12.57;$$
(3)\ If $y_1=y_2=50, y_3=0$, then $h=2$, $\sharp C_h=2$,  $g=10$ and
$$CI_h=\sqrt{\sqrt{2}\times[50+(\sqrt{2}-1)\times 50]}=10;$$
$$CI_g=\sqrt{\sqrt{10}\times[50+(\sqrt{2}-1)\times 50]}=14.95;$$
(4)\ If $y_1=100, y_{2}=0$, then   $h=1$, $\sharp C_h=1$, $g=10$, $CI_h=10$  and
$CI_g=17.78.$\\
From the calculation results above, we  see that the cases (1)-(4) have the same  $CI_h$,  but  with different $CI_g$. More precisely, we find that
$R_1\prec R_2 \prec R_3 \prec R_4$.

  \vskip 0.2cm
 \section{Concluding remarks}
\setcounter{equation}{0}

Based on the Choquet integral  and the foundation of  the $h$-index and $g$-index we have
introduced the $CI$-indices within the $h$-core and $g$-core, and then we consider  the $CI$-index in the core of all  citations.
These new indices eliminate some of the disadvantages of the $h$-index,  $g$-index, $A$-index and $R$-index and has a notable feature that
highly cited papers have highly weights and  lowly cited papers have lowly weights. After that, we propose a  new method to
compare the  academic achievements of two researchers.
This research   has not  taken into account the effect of multiple authorship as in Hirsch (2010, 2019) and the effect of  self-citation as in Bartneck and  Kokkelmans (2011) and others  which could be an excellent  direction for
further research.   We  hope that this new $CI$-index will be further studied and used in practical assessments.

\noindent{\bf Acknowledgements.} \ 
The research   was supported by the National
Natural Science Foundation of China (No. 11571198).

\end{document}